%% file: main.tex
\documentclass[10pt,conference]{IEEEtran} 
\IEEEoverridecommandlockouts

\IEEEpubid{\begin{minipage}{\textwidth}\ \\[12pt] \centering
\\ \vspace{2cm}
  \copyright 2024 IEEE. Personal use of this material is permitted.  Permission from IEEE must be obtained for all other uses, in any current or future media, including reprinting/republishing this material for advertising or promotional purposes, creating new collective works, for resale or redistribution to servers or lists, or reuse of any copyrighted component of this work in other works.
\end{minipage}}

\usepackage{wrapfig}
\usepackage{tikzit}
\input{Styles.tikzstyles}

\usepackage{cite}
\usepackage{amsmath,amssymb,amsfonts}
\usepackage{algorithmic}
\usepackage{graphicx}
\usepackage{textcomp}
\usepackage{pdflscape}

\usepackage{tabularx}
\usepackage{color}
\usepackage{amssymb,amsmath, amsthm, amsfonts,amstext,mathrsfs}

\usepackage{braket}

\definecolor{darkblue}{rgb}{0,0,0.5}
\definecolor{darkgreen}{rgb}{0,0.5,0}
\definecolor{darkred}{rgb}{0.9,0,0}
\usepackage[
    colorlinks=true,
    linkcolor=darkblue,
    urlcolor=darkblue,
    citecolor=darkgreen
]{hyperref}
\usepackage[capitalize]{cleveref}
\newtheorem{definition}{Definition}
\def\BibTeX{{\rm B\kern-.05em{\sc i\kern-.025em b}\kern-.08em
    T\kern-.1667em\lower.7ex\hbox{E}\kern-.125emX}}
    
\begin{document}
\title{A Quantum Vault Scheme for Digital Currency}

\author{\IEEEauthorblockN{ Anne Broadbent}
\IEEEauthorblockA{\textit{Dept.~of Mathematics and Statistics} \\
\textit{University of Ottawa}\\
Ottawa, Canada \\
abroadbe@uottawa.ca }
\and
\IEEEauthorblockN{ Raza Ali Kazmi}
\IEEEauthorblockA{\textit{Dept. of Corporate Services and Data} \\
\textit{Bank of Canada}\\
Ottawa, Canada \\
razaalikazmi4@gmail.com}
\and
\IEEEauthorblockN{ Cyrus Minwalla}
\IEEEauthorblockA{\textit{Dept. of Information Technology Services} \\
\textit{Bank of Canada}\\
Ottawa, Canada\\
cminwalla@bank-banque-canada.ca}}

\maketitle

\begin{abstract}
 A digital currency is money in a digital form. In this model, maintaining integrity of the supply is a core concern, therefore protections against double-spending are often at the heart of a secure digital money scheme. Quantum money exploits the quantum mechanical principle of no-cloning to enable a currency that is immune to double spending. One of the challenges of the scheme is that users require technology that is currently out of reach. Here, we propose a model for quantum currency, which alleviates the need for quantum wallets by delegating quantum storage and processing to an intermediary that we call a \emph{quantum vault}. We develop the basic building blocks of this quantum-enabled digital currency and discuss its benefits and challenges. 
\end{abstract}

\begin{IEEEkeywords}
quantum finance, quantum money, digital currency
\end{IEEEkeywords}

\section{Introduction}
The digital economy accounts for an ever-increasing share of global economic activity. These transactions settle through digital payments represented as bits in some form or another. For completeness, we consider a digital currency (DC) capable of both typical digital payments involving an intermediary as well as \emph{offline} payments, where value is transferred locally in a peer-to-peer transaction without requiring an intermediary for settlement. Built on classical computing principles, these currencies require ever-increasing layers of controls to protect against counterfeiting, unauthorized spending and fraudulent activity. Such controls can span the gamut of cryptographic techniques \cite{CFN1990}, \cite{CLB07}, trusted hardware \cite{Visa20} \cite{GFRSV20patent} and consensus protocols \cite{Nakamoto08} \cite{HotStuff19}  to protect the currency.  

Despite best efforts, security controls can only minimize, not eliminate the double-spending threat, the source of which is the fact that transmission and subsequent deletion of digital assets are two distinct steps that cannot be made atomic in a classical computer. In the event that a payer (Alice) wishes to transfer funds from her device to the device of a payee/receiver (Bob), such a move between two devices is never atomic but consists of a copy instruction followed by a delete instruction. It is possible that a malicious third party (Alice), with sufficient effort and diligence, can always find a way to circumvent the delete instruction.  

The use of quantum information in DC mitigates the double-spending issue by leveraging the no-cloning principle that prohibits the copy of an arbitrary quantum state\cite{Par70,WZ82} . This property is generally useful for all payments and particularly relevant for offline payments where an intermediary may not be available to attest to the truth. The solution, however, introduces the requirement of quantum wallets where the technology is challenging and costly in regards of funds storage and processing integrity.

\subsection{Contributions and Outline}
We propose a model for a \emph{quantum} currency (QC) that achieves a reasonable trade-off towards solving the above two issues. Our model involves an intermediary \emph{quantum vault}, which is a quantum-enabled Money Services Business ({MSB}), to which wallets (end-users) delegate the storage and processing of quantum money. This model satisfies all of the properties of digital currency and is a clear improvement over existing currency schemes based on classical assumptions.  

The remainder of the paper is structured as follows:  \cref{sec:Background} presents a literature review of digital currency, with a deeper dive into existing work involving quantum wallets (\cref{sec:Scheme with Quantum Wallet}). This is followed by a description of the proposed quantum vault scheme (\cref{sec:Classical Wallets}), the security and privacy properties of which are discussed in \cref{sec:Security and Privacy Analysis}, ending with concluding thoughts and avenues for future work.

\section{Background}
\label{sec:Background}

\subsection{Properties of Digital Currencies}
A digital currency is money in a digital form that can be used to store value and make electronic payments. Digital currencies may be private, such as money held at financial institutions and crypto-currencies, or public, such as that issued by a central bank. Presented as follows are selected properties that various forms of digital currency should satisfy. 

\subsubsection{Authenticity}
A holder of a unit of funds can prove that the funds originated from the entity authorized to issue funds. In classical systems, this typically requires validating a signature generated by a central authority or a root certificate, or confirming the token's existence on a publicly verifiable source, such as a blockchain. 

\subsubsection{Double-spend resistance}
A unit of funds cannot be spent in two or more transactions without a change in ownership occurring between transactions. The true owner of funds can only spend the funds once, after which they are deleted or altered in a manner that prevent further attempts to spend again. 

\subsubsection{Transitivity}
Pertaining specifically to extended offline digital currencies, this property ensures that funds can be transferred multiple times in a bilateral (offline) fashion without requiring a connection to a third-party for settlement or validation. Local parties must be able to satisfy the other properties of provenance, independence, and counterfeit detection without requiring assistance from third parties. 

\subsubsection{Independence}
A property whereby a unit of funds is distinct and independent of all other units of funds in the ecosystem. This property implies that multiple units of funds can be transacted without depending on the outcome of transactions of other units of funds. It is notable that many crypto-currencies based on blockchain technology experience performance bottlenecks since this requirement is not satisfied.

\subsubsection{Confidentiality}
A property where information about the transaction is only available to the parties required to settle the transaction and those required to perform compliance on the transaction. Recorded transaction details are minimized, similarly, transaction amounts and histories are protected from disclosure to third parties except for those authorized to access it.  

\subsubsection{Offline functionality}
The ability to make a bilateral transaction between a payor and a payee without requiring network connectivity or a third-party at the time of transaction. Settlement can be deferred in the case of intermittent offline, or immediate, in the case of extended offline \cite{MMSLH23}. Not all forms of digital money can satisfy this requirement.

\subsection{Private-Key Quantum Money}
\label{sec:private-key-quantum money}

In the late 1960s, Wiesner~\cite{Wie83} had the visionary idea that quantum information could be used to create unforgeable bank notes (according to~\cite{BBB14}, Wiesner’s original manuscript was written in 1968, but not published until 1983). In modern terminology, Wiesner's concept is called \emph{Private-Key quantum money}. In this section, we survey Wiesner's proposal and related work. We note that much of this section is from a survey on \emph{quantum cryptography beyond quantum key distribution}, which is a contribution by one of the authors of the current document~\cite{BS16}.

\subsubsection{Conjugate Coding}
Conjugate coding is based on the principle that classical information can be encoded into conjugate quantum bases. This primitive is extremely important in quantum cryptography—--in fact, the vast majority of quantum cryptographic protocols (including the famous BB84 quantum key distribution~\cite{BB84}) exploit conjugate coding in some form or another. 

The principle of conjugate coding is straightforward. For clarity of presentation and consistency with commonly used terminology, we associate a qubit with a photon (a particle of light), and use photon polarization as a quantum degree of freedom. Among others, photons can be polarized horizontally, vertically, diagonally to the right, or diagonally to the left. Photon polarization is a quantum property, and by associating horizontal polarization to $\ket{0}$, vertical to~$\ket{1}$, diagonal right to $\ket{+} : = \frac{1}{\sqrt{2}}(\ket{0} + \ket{1}) $ and  diagonal left to $\ket{-}:=\frac{1}{\sqrt{2}}(\ket{0} - \ket{1})$, we can define two \emph{mutually unbiased} (\emph{conjugate}) bases as $B_1:=\{\ket{0}, \ket{1}\}, B_2:=\{\ket{+}, \ket{-}\}$, where we refer to $B_1$ as the \emph{computational} basis and $B_2$ as the \emph{diagonal} basis.  

The relevance of conjugate coding to cryptography is summarized by two key features that were mentioned and exploited in Wiesner’s work:
\begin{enumerate}
    \item  Measuring in one basis irrevocably destroys any information about the encoding in its conjugate basis.
    \item The originator of the quantum encoding can verify its authenticity by measuring in the known encoding basis; however, without knowledge of the encoding basis, and given access to a single encoded state, no third party can create two quantum states that pass this verification procedure with high probability.
\end{enumerate}

To explain the first property, recall the well-known Heisenberg uncertainty principle~\cite{Hei27}, which forbids learning both the position and momentum of a particle precisely and simultaneously. In terms of photon polarization, and for a single photon, let us denote by $P_X$ the distribution of outcomes when measuring the photon in the computation basis and by $Q_X$ the distribution of outcomes when measuring the photon in the diagonal basis.   Maassen and Uffink~\cite{MU88} showed an uncertainty relation: $H(P_X) + H(Q_X) \geq 1$, where $H$ is the \emph{Shannon entropy}, and information-theoretic measure of uncertainty. 
Intuitively, such a relation quantifies the fact that one can know the outcome exactly in one basis, but consequently has complete uncertainty in the other basis.

\subsubsection{Wiesner Quantum Money}

Wiesner’s proposal consists of quantum banknotes created by encoding quantum particles using conjugate coding, with both the classical information and basis choice being chosen as random bitstrings. Thus, a banknote is comprised of a sequence of $n$ single qubits, chosen randomly from the states $\{\ket{0}, \ket{1}, \ket{+}, \ket{-}\}$. As discussed above, the originator of the quantum banknote (typically called “the bank”) can verify that a quantum banknote is genuine, yet quantum mechanics prevents any possibility of counterfeiting. Clearly, such functionality is beyond what classical physics can offer. Since any digital record can be copied, classical information cannot be used for uncloneability, and not even computational assumptions will help in this regard.

\subsubsection{Proof of security for Wiesner's scheme}
The first proof of security for Wiesner's scheme appeared 30 years after the publication of the scheme, and is based on semi-definite programming~\cite{MVW13}. The result formally postulates a \emph{counterfeiting attack}, in which the bank issues an authentic bank note (consisting in $n$ qubits as give above), which is then given directly to a counterfeiter, who then creates two $n$-qubit systems, both of which are then verified using the bank's verification procedure as given above. The result of~\cite{MVW13} is that the \emph{optimal counterfeiting probability} (the probability that the bank accepts \emph{both} $n$-qubit systems as valid) is $\left({\frac{3}{4}}\right)^n$. We note that~\cite{MVW13} also show that this result is tight by giving an explicit optimal attack that reaches this bound. 

\subsubsection{Extensions to Wiesner}
\label{sec:private-key-related work}

Wiesner’s work was improved and extended, and its limitations were also studied: 

\paragraph{Returning state after verification}  Variants of Wiesner’s scheme in which quantum encodings are returned after validation were studied: In all cases (whether the post-verification state is always returned \cite{FGH+10,Lut10arxiv}, or the post-verification state is returned only for encodings that are deemed valid~\cite{BNSU14arxiv}, the resulting protocol was shown to be insecure. 
\paragraph{Noise-tolerance}
A noise-tolerant version of Wiesner's scheme was developed: \cite{PYJ+12}. This is particularly relevant for experimental demonstrations, 
and further refined to a nearly optimal scheme in~\cite{AA17}.
\paragraph{Classical interaction only} 
Further work has studied the possibility of private-key quantum money that can be verified using only classical interaction with the bank \cite{Gav12, MVW13}.
\paragraph{Full anonymity} Quantum coins were proposed \cite{MS10}, where the anonymity of coins is emulated.

\subsection{Public-Key Quantum Money}
\label{sec:Public-key Quantum money}

The Wiesner scheme and extensions relying on a private-key quantum money construction have a major drawback in that \emph{only the originator of the banknote can verify its validity}. Indeed, the information required to validate the banknote is the \emph{same} information that can be used to make a fresh copy of the banknote. It is therefore impossible for general users of any private-key money system to be able to verify the banknotes. 

The aforementioned also implies that offline transactions are impossible since the originator must be an intermediary in each transaction. In addition, the communication of the currency must be through a quantum channel as opposed to the classical channel. By separating the procedure of minting and verification, it is possible to envisage what is called \emph{public-key quantum money}, which we define next. 

\paragraph{Notation} We use below the following conventions: 
\begin{enumerate}
\item PPT: Probabilistic Polynomial Time (\emph{i.e.}, an efficient classical process)
\item QPT: Quantum Polynomial Time (\emph{i.e.}, an efficient quantum process)
\end{enumerate}

\begin{definition}
{\em A Public-key quantum money scheme} is a tuple of 4 algorithms  ({\tt Gen}, {\tt BankMint}, {\tt RecMint}, {\tt QV})
 \begin{enumerate}
     \item The {\tt Gen} is a classical PPT  algorithm that takes  a security parameter $\lambda$ as input and outputs a public/secret key pair $(pk,sk)\rightarrow \tt{Gen(1^\lambda)}.$
     
     \item $|\$\rangle\leftarrow \langle{\tt BankMint}(sk), {\tt RecMint}(pk)\rangle$ a classical two-party interactive protocol between a classical PPT algorithm {\tt BankMint} and a QPT algorithm {\tt RecMint}. At the end of the interaction, the receiver has a quantum banknote~$|\$\rangle$.
    
     \item $(b,|\$^\prime\rangle) \leftarrow {\tt QV}(pk,|\$\rangle$): A QPT verification algorithm that takes  as input the public key~$pk$ and a candidate banknote $|\$\rangle$ and outputs a banknote $|\$^\prime\rangle$ along with a bit $b\in\{0,1\}$ indicating valid or invalid respectively.
   
     \end{enumerate}

In addition, a public-key quantum money scheme may provide an additional feature called \emph{Classical Certificates of Destruction (CCoD)}~\cite{RS22,Shm22} that provides the additional two algorithms  ({\tt GenCert}, {\tt CV})

\begin{enumerate}  
     \item[4.] ${\tt crt}\leftarrow{\tt GenCert}(pk,|\$\rangle)$: A QPT algorithm that receives as input the public key $pk$ and a candidate banknote $|\$\rangle$ and outputs a classical string {\tt crt}. This allows the sender to destroy a banknote and produces a classical certificate of destruction. 
     \item[5.] ${\tt CV}(pk,{\tt crt})\in\{0,1\}$: A classical algorithm that takes as input the public key $pk$ and a classical string {\tt crt}, and outputs a bit indicating verification success or failure.
\end{enumerate}

\end{definition}

The above formal interfaces can be matched with formal \emph{correctness} and \emph{security} definition; we refer to~\cite{Shm22} for details and summarize the intuition below. 

\paragraph{Correctness}--- ``Honest banknotes are accepted'': If we first run {\tt Gen}, followed by $\langle{\tt BankMint}(sk), {\tt RecMint}(pk)\rangle$, and then feed the output into {\tt QV}, then the outcome is $b=1$ with overwhelming probability.

\paragraph{Security against counterfeiting}--- ``Protecting  the bank'': Given one valid banknote, an adversary cannot output both a quantum banknote and a corresponding valid classical certificate of destruction for it. This mutual exclusivity is a powerful complement to no-cloning, as it guarantees that once a token is spent it is irrevocably destroyed in an attestable manner; a property not possible to achieve in a classical representation of money.

\paragraph{Security against sabotage} ---``Protecting wallets in the system'': Given that wallets hold quantum banknotes, when a wallet is given a quantum banknote that  passes the public quantum verification ${\tt QV}(pk, \cdot)$ once, it is guaranteed that the banknote will pass all further quantum verifications with overwhelming probability. Once a banknote is ready to be redeemed, it can be destroyed with $\tt{GenCert}(pk, \cdot)$ generating a valid classical certificate of destruction {\tt crt} in the process, which is verifiable by $\tt{CV}(pk, \cdot)$.

\paragraph{Classical Minting} ---The definitions outlined herein specify a \emph{classical} minting procedure by which a receiver constructs a quantum state using classical interaction (only) with the bank. If this property is not satisfied, such a procedure is then called a \emph{quantum minting} algorithm.

\subsubsection{Public-key quantum money}
Early work of Bennett, Brassard, Breidbart and Wiesner \cite{BBBW82} illustrated how computational assumptions can be combined with conjugate coding to achieve an early type of public verifiability for the encoded states. They coined their invention \emph{unforgeable subway tokens}. This early proposal came without a security reduction, and was intuitively based on the idea that the factoring problem is hard. 

\subsubsection{Knot-based public-key quantum money}
Farhi, Gosset, Hassidim, Lutomirski and Shor present a public-key quantum money construction that is locally verifiable on a quantum device \cite{FGH+10}. The scheme requires a fully quantum communication channel established between the issuing bank and the receiver during the minting process. The security is based on knot theory conjectures that are not widely studied. 
 
\subsubsection{Hidden Subspaces}
Aaronson and Christiano\cite{AC12} developed a public-key quantum money scheme that built on linear algebraic principles: A money state is an $n$-qubit state that is a superposition of all $n$-bit strings in an $n/2$-dimensional random subspace~$A$ of the $n$-bit strings. 

Verification of such a quantum money state is akin to verifying that the state is in the span of the defined subspace, and that its Fourier transform is in the span of the orthogonal subspace, $A^\perp$. Security is proven under the assumption that these verification mechanisms have access to appropriate \emph{oracles}. The \emph{conjecture} is then that the verification can be given in an instantiation that would be in an obfuscated form (regardless of mechanism), and that security would still hold. 

However, no such secure instantiations are currently known, and some prior proposals have since been broken~\cite{LAF+10,CDF+19}.

\subsubsection{Quantum Lightning} 
\label{sec:Lightning}
The first proposal for public-key semi-quantum (where the minting is classical) is based on a concept called \emph{quantum lightning} \cite{Zha19}, which is essentially a non-interactive and reusable classical delegation of sampling states that are uncloneable and publicly verifiable. In particular, Quantum Lightning gives a solution to the classical minting problem of public-key quantum money (but does
not necessarily provide classical proofs of destruction of banknotes).
Zhandry gave a construction of Quantum Lightning based on
a new computational assumption. The security of Zhandry’s construction
was later called into question when Roberts showed that
the computational assumption is broken~\cite{Rob21}.

However, one version of Zhandry's scheme still stands, which is the one that relies on the security of \emph{quantum-secure indistinguishability obfuscation}. Such a scheme is a compiler, $i\mathcal{O}$ that takes as input a circuit and outputs another, functionally equivalent circuit, with the property that for two circuits $\mathcal{C}_1, \mathcal{C}_2$ that are functionally equivalent, their obfuscations $i\mathcal{O}(\mathcal{C}_1)$, $i\mathcal{O}(\mathcal{C}_2)$ are computationally indistinguishable.    Currently, the post-quantum security of iO remains poorly understood, with all known constructions of quantum-secure iO \cite{GGH15,BGMZ18, BDGM20eprint,WW21}  being at best labeled as candidates.

\subsubsection{Quantum Money from Lattices}
Khesin,  Lu, and Shor proposed a scheme for public-key quantum money using Gaussian superpositions over random lattices \cite{KLS22arxiv}. Although the security was based on the  hardness of the short vector problem from lattice-based cryptography, it was not formally reduced to a well-studied problem, and was recently broken~\cite{LMZ23}.

\subsubsection{Public-Key \emph{semi-quantum} money}
A central question related to public-key quantum money is whether or not the minting process can be a classical algorithm. In particular, such a scheme relies on local quantum computation and only classical communication. 

Radian and Sattath~\cite{RS22} proposed a scheme referred to as public-key \emph{semi-quantum} money. They demonstrated that a semi-quantum money scheme can be achieved based on a scheme with classical minting and with 
CCoDs. Intuitively, this is possible given that any quantum wallet can return a currently held, valid quantum banknote to the classical bank. Specifically, a quantum wallet can generate a classical certificate {\tt crt} for its quantum banknote, which guarantees that the quantum state has been destroyed and subsequently cannot pass public quantum verification. This means that when the bank receives a valid{\tt crt}, it can safely re-issue one or more banknotes of equivalent value to the intended parties.  Note however that this type of transaction must be performed online with the bank; quantum communication is required to perform purely offline transactions between quantum wallets.

The state-of-the art in semi-quantum money (with classical minting and with CCoDs) is work by Shmueli \cite{Shm22}, subsequently referred as SRS (Shmueli, Radian, Sattath), which proposes a scheme with security based on the following two conditions: 
\begin{enumerate}
\item quantum-secure indistinguishability obfuscation (iO); and 
\item the sub-exponential hardness of the Learning With Errors (LWE) problem.
\end{enumerate}

The technical centerpiece is a new 3-message protocol, where a classical computer can delegate to a quantum computer the generation of a quantum state that is both unclonable and publicly verifiable. The main technical tools are Quantum Fully Homomorphic Encryption (QFHE) and iO. Both of these primitives are topics of current study (both in terms of practicality and security), with candidates \cite{Mah18b, Bra18} for QFHE and \cite{GGH15,BGMZ18, BDGM20eprint,WW21} for iO.

\section{Scheme with A Quantum Wallet}
\label{sec:Scheme with Quantum Wallet}

We further elaborate on the SRS construction in this section. The issuing authority, equipped with a classical computer, delegates to the wallet, equipped with a quantum computer, the task of generating, storing and processing a quantum state that defines the banknote. The resultant state is publicly verifiable and impervious to cloning. All communication between the issuing authority and the wallet (i.e., banknote minting, verification of banknote destruction)  occurs through a classical communication infrastructure. This setup results in a system where the sole quantum communication and computation is confined to and shared between wallets.

\subsection{Entities and Roles}
\paragraph{Issuing Authority}--- A classical digital trusted entity tasked with issuing the classical component of a banknote upon request from a Wallet engaged in the acquisition of banknotes or online payment transactions (i.e., transactions with the involvement of the Bank). The Wallet entity is entrusted with the quantum minting aspect of the banknote. 

\paragraph{Wallet (end user)}--- A quantum digital entity designed to safeguard an individual's quantum banknotes and facilitate payment transactions for buying or selling goods/services. Its duties encompass quantum minting (as a recipient), overseeing and storing quantum banknotes, such as acquiring or disposing of them, as well as managing online or offline payment transactions. 

\subsection{Processes}

\subsubsection{On-Demand Banknote Minting (acquisition)} This process involves an individual seeking to obtain a new banknote, which adopts a digital format encapsulating a value, a unique identifier denoting its origin and legitimacy (i.e., the public key), and a series of anti-counterfeit markings (i.e., the classical and quantum cipher banknotes using the secret key). The process unfolds through three sequential interactions between the individual's Wallet and the Issuing Authority: \textbf{classical banknote minting}, \textbf{quantum banknote minting}, and \textbf{banknote validation}. 

\begin{enumerate}
    \item Classical Banknote Minting: Upon the Wallet's request, the Issuing Authority generates the classical banknote using the \textit{BankMint algorithm}, taking a public/secret key pair as input (i.e., the \textit{Gen algorithm}). The Issuing Authority then transfers the classical banknote, along with the public key, to the Wallet. 

    \item  Quantum Banknote Minting: After receiving the Issuing Authority's classical banknote and public key, the Wallet mints the Quantum banknote using the \textit{RecMint algorithm}. This results in a quantum state and a classical ciphertext. The Wallet then transfers the classical ciphertext back to the Issuing Authority.

    \item Banknote Validation: Upon receiving the Wallet's classical ciphertext, the Issuing Authority decrypts it and verifies its match with the original ciphertext generated by the \textit{BankMint algorithm}. This information is then used to compute the final public key, serving as a unique identifier for provenance, traceability, and legitimacy purposes. The Issuing Authority transfers this final public key back to the Wallet.
\end{enumerate}
At the conclusion of this process, the Issuing Authority retains a classical banknote associated with a public key, while the Wallet stores the equivalent quantum banknote linked to the same public key.

\subsubsection{Offline Payment Transaction} This process occurs when an individual (Payer) intends to transfer money (a banknote) to another individual (Receiver) in exchange for a purchased good or service. It entails a two-step interaction process exclusively involving the Wallets: \textbf{banknote transfer} and \textbf{banknote validation}. Notably, the Issuing Authority does not play a role in the transaction.

\begin{enumerate}
\item Banknote Transfer: The Wallet (Payer) physically hands over the banknote, along with its public key, to the Wallet (Receiver). It is essential to emphasize that the transfer process does not involve the bit-wise duplication of a banknote, as seen in the classical world (e.g., copy then delete). The quantum mechanism prevents any duplication of the banknote and ensures immunity to double spending.

\item Banknote Validation: The Wallet (Receiver) verifies the authenticity of the received banknote using the \textit{QV algorithm}. If the banknote is deemed invalid, the Wallet (Receiver) discards it and notifies the Wallet (Payer) of its destruction.
\end{enumerate}

At the conclusion of this process, the Wallet (Receiver) either retains the banknote along with its public key or discards it if it is found to be invalid. In the latter scenario, another payment transaction must occur between the Payer and the Receiver, although this falls outside the scope of the current process. It is noted that there is a potential trust issue if the recipient can unilaterally decide to discard; see \cref{sec:Conclusions and Future Work} for possible future work.

\subsubsection{Online Payment Transaction} In this process, an individual (Payer) intends to transfer money (a banknote) to another individual (Receiver) in exchange for a purchased good or service. However, the Issuing Authority aims to maintain strict control over the total funds in circulation to prevent banknote counterfeiting and duplication, ensuring immunity against double-spending. The process involves three interactions between the Wallets (Payer, Receiver) and the Issuing Authority: \textbf{banknote destruction}, \textbf{banknote destruction confirmation}, and \textbf{banknote minting}.
\begin{enumerate}

\item Banknote Destruction: The Wallet (Payer) utilizes the \textit{GenCert algorithm} to destroy the banknote, resulting in a classical certificate of destruction. This certificate is then transferred to the Wallet (Receiver). It is important to note that the quantum banknote is physically deleted from the Wallet (Payer).

 \item Banknote Destruction Confirmation: The Wallet (Receiver) requests the Issuing Authority to mint a new banknote with an equivalent value to the destroyed one. Along with this request, the Wallet (Receiver) sends the classical certificate of destruction of the original banknote. The Issuing Authority validates and confirms that the received certificate of destruction corresponds to the original banknote using the \textit{CV algorithm}.

 \item Banknote Minting: Upon successful validation of the certificate of destruction, the Issuing Authority and the Wallet (Receiver) proceed to the \textit{banknote minting (acquisition) process}, as outlined above.
\end{enumerate}
 At the conclusion of this process, the Issuing Authority has minted a new banknote with a value equivalent to the originally destroyed banknote, the Wallet (Receiver) has acquired the new banknote, and the Wallet (Payer) has disposed of its original banknote.
\section{Making Wallets Classical}
\label{sec:Classical Wallets}

Herein we propose a classical wallet scheme that improves upon some of the shortcomings in SRS. To participate in SRS, users require portable quantum wallets that are able to establish secure quantum communications channels. Given that the technological state-of-the-art required to achieve this level of functionality is likely decades into the future, a more practical custodial model where funds are held in a quantum state at intermediaries and transferred over quantum channels established between intermediaries is preferred. To that end, an intermediary \emph{quantum vault}, designated as a quantum-enabled Money Services Business (MSB), is included in the system. Wallets, representing end-users, delegate tasks such as interactions with the issuing authority, as well as the creation, storage, and processing of quantum banknotes, to the intermediary (and its respective quantum vault) associated with their Wallet. Consequently, the Wallet can now operate as software on a classical system and authenticates with the intermediary using classical means. This configuration establishes a system wherein the exclusive infrastructure for quantum communication and computing is confined to and shared between MSBs, obviating the need for end-users to carry portable quantum wallets.

\subsection{Quantum Vault System}
\begin{figure*}[t]
    \begin{center}
        \scalebox{1.0}{\input{QuantumVaultSystem.tikz}}
        \caption{Quantum Vault System}
        \label{fig:QVS}
    \end{center}
\end{figure*}
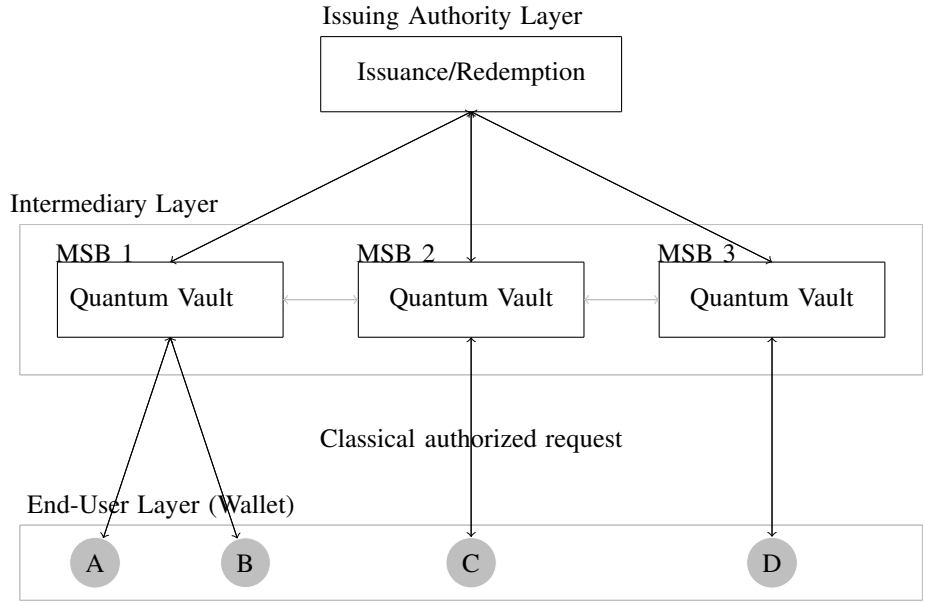

A quantum vault system consists of three layers: the issuing authority classical layer responsible for minting the classical banknotes, the quantum intermediary layer composed of quantum vaults or MSBs responsible for minting and storing the quantum banknotes, as well as managing the payment transactions, and the end-user classical layer composed of wallets whose main task is to handle end-user (i.e., people) interactions to initiate banknote acquisition and payment transactions (see~\cref{fig:QVS}).

\subsection{Entities and Roles}

\paragraph{Issuing Authority}---A classical digital trusted entity tasked with issuing the classical component of a banknote upon request from an MSB engaged in the acquisition of banknotes or online payment transactions (i.e., transactions with the involvement of the Bank). The MSB entity is entrusted with the quantum minting aspect of the banknote when and as instructed by the Issuing Authority. 
\paragraph{MSB -- Money Services Business}--- A quantum digital entity designed to safeguard individual Wallet's quantum banknotes and facilitate payment transactions for buying or selling goods/services. Its duties encompass quantum minting, overseeing and storing quantum banknotes (i.e., acquiring or disposing of them), as well as managing online payment transactions between two individual end-users that are intra- and inter-MSB. The MSB can be classified as a `custodian,' indicating its complete responsibility for the management of the Wallet's (end users) banknotes.

\paragraph{Wallet (end user)}--- A classical digital entity possessing the sole capability of issuing commands from an individual (e.g., acquire banknotes, initiate transactions) to the Money Services Business and engaging in communication with another individual's Wallet regarding the agreement on payment transactions. 

\subsection{Processes}

\subsubsection{On-Demand Banknote Minting (acquisition)} 
\begin{figure*}[t]
    \begin{center}
        \scalebox{0.9}{\input{OnDemandBanknoteMinting.tikz}
        }
        \caption{On-Demand Banknote Minting}
        \label{fig:ODM}
    \end{center}
\end{figure*}
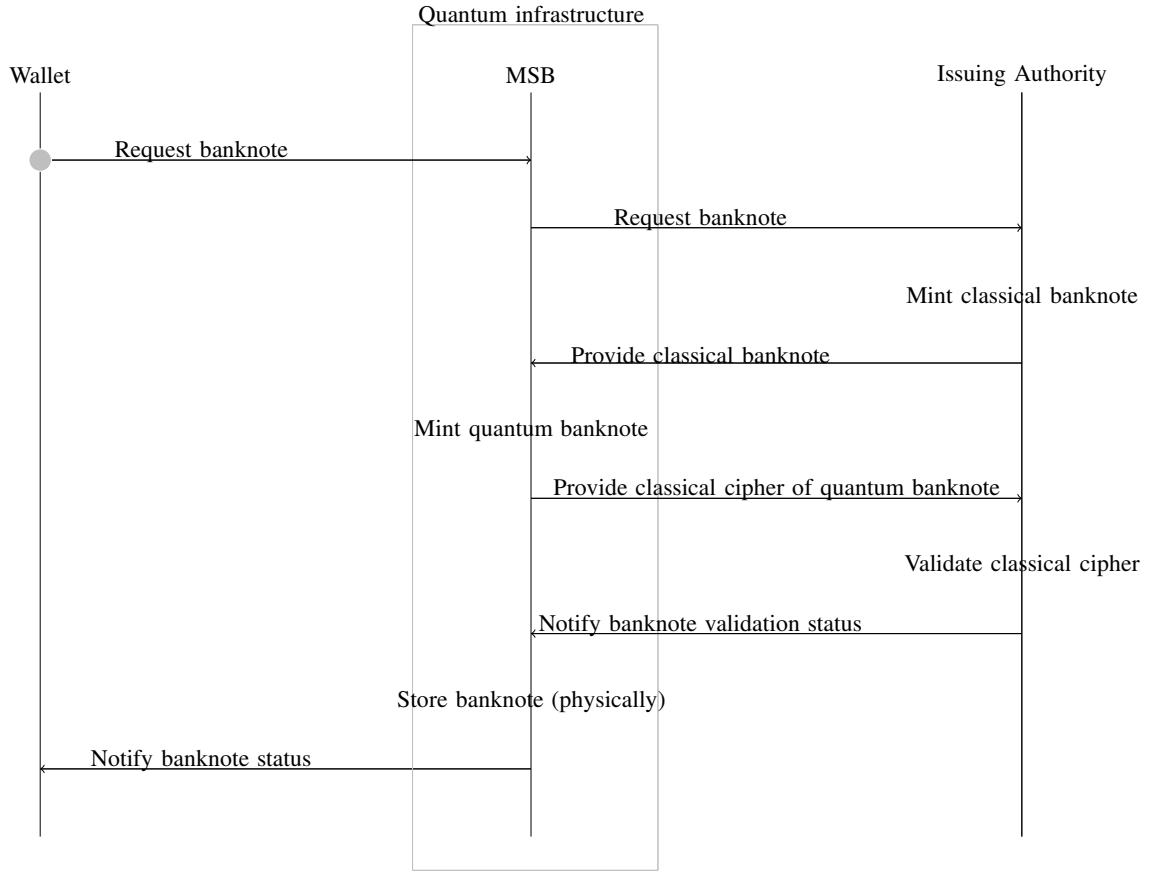

This process involves an individual seeking to obtain a new banknote, which adopts a digital format encapsulating a value, a unique identifier denoting its origin and legitimacy (i.e., the public key), and a series of anti-counterfeit markings (i.e., the classical and quantum cipher banknotes using the secret key). The process unfolds through three sequential interactions between the individual's Wallet, the MSB and the Issuing Authority: \textbf{classical banknote minting, quantum banknote minting,} and \textbf{banknote validation} (see~\cref{fig:ODM}).
\begin{enumerate}
\item Classical Banknote Minting: Upon the MSB's request originally triggered by the Wallet, the tIssuing Authority generates the classical banknote using the \textit{BankMint algorithm}, taking a public/secret key pair as input (i.e., the \textit{Gen algorithm}). The Issuing Authority then transfers the classical banknote, along with the public key, to the~MSB.

\item Quantum Banknote Minting: After receiving the Issuing Authority's classical banknote and public key, the MSB mints the Quantum banknote using the \textit{RecMint algorithm}. This results in a quantum state and a classical ciphertext. The MSB then transfers the classical ciphertext back to the Issuing Authority.

\item Banknote Validation: Upon receiving the Wallet's classical ciphertext, the Issuing Authority decrypts it and verifies its match with the original ciphertext generated by the \textit{BankMint algorithm}. This information is then used to compute the final public key, serving as a unique identifier for provenance, traceability, and legitimacy purposes. The Issuing Authority transfers this final public key back to the MSB.
\end{enumerate}

At the conclusion of this process, the Issuing Authority retains a classical banknote associated with a public key, while the MSB stores the equivalent quantum banknote linked to the same public key.

\subsubsection{Quantum Banknote Transfer}  
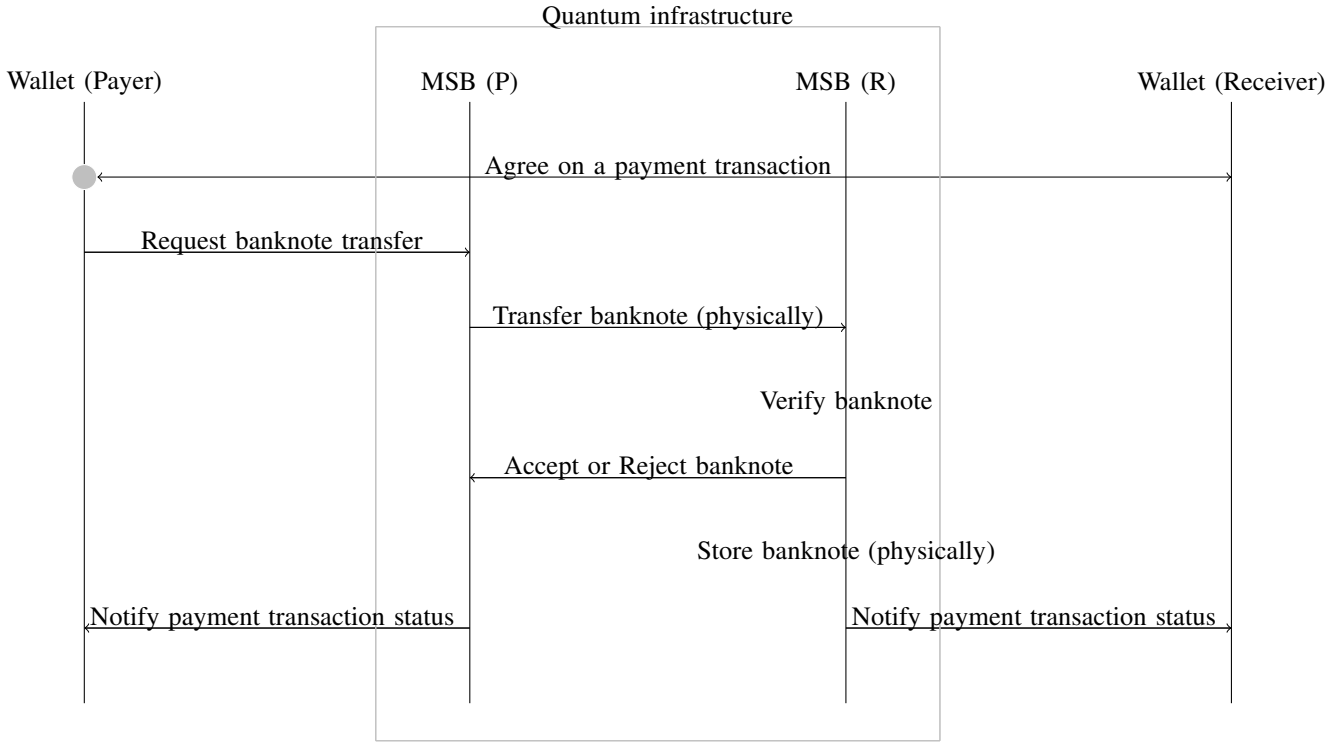
\begin{figure*}[t]
    \begin{center}
        \scalebox{1.0}{\input{QuantumBanknoteTransferInterMSB.tikz}}
        \caption{Quantum Banknote Transfer - Inter MSB}
        \label{fig:QBT}
    \end{center}
\end{figure*}

This process occurs when an individual (Payer) intends to transfer money (a banknote) to another individual (Receiver) in exchange for a purchased good or service. It entails a three-step interaction process involving the Wallets and their custodial MSB only: \textbf{payment transaction agreement, banknote transfer} and \textbf{banknote validation}. Notably, the Issuing Authority does not play a role in the transaction but it requires the intervention of a third-party (the MSB) at the time of transaction. This also implies an additional network connectivity beyond local communications between the payer and receiver (payee) to complete the transfer of the banknote (see~\cref{fig:QBT}).

For ease of reading, the MSB overseeing the Wallet (Payer) is denoted as MSB (P), while the MSB managing the Wallet (Receiver) is referred to as MSB (R).

\begin{enumerate}
    \item Payment Transaction Agreement: the Wallet (Payer) and the Wallet (Receiver) agree upon a payment transaction (value).
    \item Banknote Transfer: Upon the Wallet (Payer) request, the MSB (P) transfers the banknote, along with its public key, to the MSB (R). 
    \item Banknote Validation: The MSB (R) verifies the authenticity of the received banknote using the \textit{QV algorithm}. If the banknote is deemed invalid, the MSB (R) discards it and notifies the MSB (P) of its destruction.
\end{enumerate}
At the conclusion of this process, the MSB (R) either retains the banknote along with its public key or discards it if it is found to be invalid. 
In the latter scenario, another payment transaction must occur between the Payer and the Receiver, although this falls outside the scope of the current process.

\subsubsection{Online Payment Transaction} 
\begin{figure*}[t]
    \begin{center}
        \scalebox{0.65}{\input{OnlinePaymentTransactionInterMSB.tikz}}
        \caption{Online Payment Transaction - Inter MSB\\ \textcolor{red}}
        \label{fig:OPT}
    \end{center}
\end{figure*}
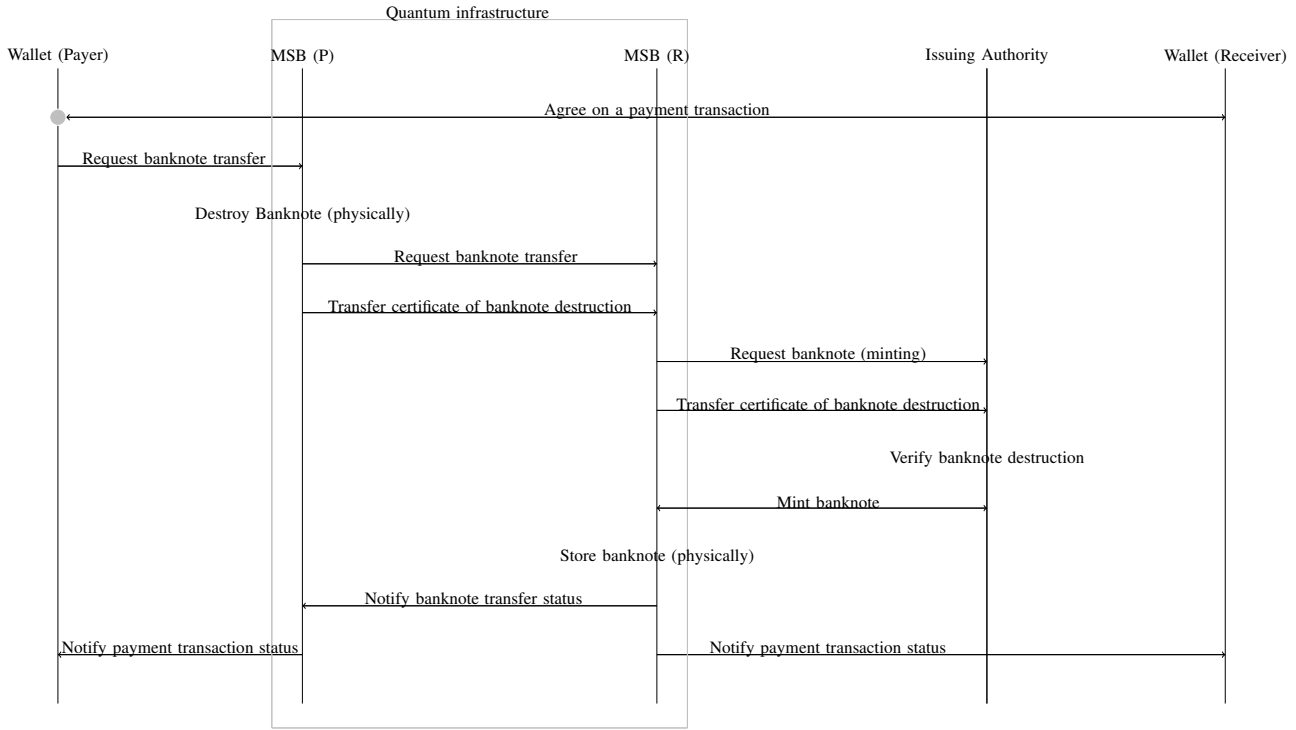

This process serves the same purpose as an offline payment transaction, where an individual (Payer) intends to transfer money (a banknote) to another individual (Receiver) in exchange for a purchased good or service. However, the Issuing Authority aims to maintain strict control over the total funds in circulation to prevent banknote counterfeiting and duplication, ensuring immunity against double-spending. The process involves four interactions between the Wallets (Payer, Receiver), their custodial MSB and the Issuing Authority: \textbf{payment transaction agreement, banknote destruction, banknote destruction confirmation,} and \textbf{banknote minting} (see~\cref{fig:OPT}).

\begin{enumerate}
\item Payment Transaction Agreement: the Wallet (Payer) and the Wallet (Receiver) agree upon a payment transaction (value).

\item Banknote Destruction: Upon the Wallet (Payer) request, the MSB (P) utilizes the \textit{GenCert algorithm} to destroy the banknote, resulting in a classical certificate of destruction. This certificate is then transferred to the MSB (R). It is important to note that the quantum banknote is physically deleted from the MSB (P).

\item Banknote Destruction Confirmation: The MSB (R) requests the Issuing Authority to mint a new banknote with an equivalent value to the destroyed one. Along with this request, the MSB (R) sends the classical certificate of destruction for the original banknote. The Issuing Authority validates and confirms that the received certificate of destruction corresponds to the original banknote using the \textit{CV algorithm}.

\item Banknote Minting: Upon successful validation of the certificate of destruction, the Issuing Authority and the MSB (R) proceed to the \textit{banknote minting (acquisition) process}, as outlined above.
\end{enumerate}

At the conclusion of this process, the Issuing Authority has minted a new banknote with a value equivalent to the originally destroyed banknote, the MSB (R) has acquired the new banknote, and the MSB (P) has disposed of its original banknote.\looseness=-1

\subsection{Communication Infrastructure and Digital Authentication}
\label{sec:Communication Infrastructure and Digital Authentication}
Our proposal builds upon a quantum infrastructure exclusively within the MSB (i.e., the MSBs are quantum devices, and the network connecting them is quantum). This approach successfully addresses the technical and costly challenges associated with implementing quantum wallets (primarily in storage) and establishing a fully developed and widely accessible quantum network (such as the Quantum Internet, which is not expected to materialize for the next 10 to 15 years).

Our proposal, however, introduces a new challenge: the authentication of wallet-MSB interacting in a classical world. Indeed, how can an MSB be assured that the wallet it engages with is authentic, and vice versa? This challenge also raises classical concerns about privacy, counterfeiting, duplication, and the unauthorized copying of wallet credentials. Furthermore, in light of the emergence of quantum computing, credential/authentication mechanisms must be fortified against quantum computing attacks that pose a threat to classical cryptographic primitives based on factorization or discrete logarithms.
To address these challenges, we propose the utilization of an anonymous credential scheme to protect user privacy\cite{CL04}.

\section{Security and Privacy Analysis}
\label{sec:Security and Privacy Analysis}

This section outlines how the proposed scheme preserves the security and privacy properties of digital currency.

\subsection{Authenticity}
At any given time, a person (the physical owner of the Wallet) can verify the authenticity of a held banknote by initiating a validation procedure through their Wallet. This involves utilizing the public key  associated with the banknote as input. The Wallet engages with the Money Service Business (MSB) to confirm the banknote's authenticity (displaying a valid/invalid status) and, by extension, to indirectly ascertain its origin as one issued by the Issuing Authority. The MSB employs the \textit{QV algorithm} to execute this verification process.

\subsection{Double-spend protection}
A key advantage of the proposed approach is that all of the properties associated with the quantum representation of money are retained despite the wallet being classical. In a quantum banknote transfer scenario, the physical quantum transfer of the banknote from the payer to the payee (receiver) ensures new and unique ownership of the banknote, a result of the no-cloning principle\cite{Par70,WZ82} inherent in quantum information, preventing the payer from retaining a record of the banknote after its transfer to the payee. 

In the context of an online payment transaction, the assurance of mutual exclusivity in the quantum representation of money is upheld. Adversaries are unable to produce both a quantum banknote and a corresponding valid classical certificate of destruction. It is important to note that the banknote in circulation through Money Service Business entities (MSBs) is entirely quantum, while its classical counterpart is securely held by the Issuing Authority and remains outside of circulation.
Ergo, it is impossible to double-spend in the proposed scheme, unlike Bitcoin and other crypto-currencies, where it is difficult, but not impossible.

\subsection{Independence}
The \textit{banknote minting (acquisition) process} ensures the uniqueness and independence of each banknote. This is achieved by constructing each banknote from random inputs, with no correlation to previously minted banknotes. The inputs include the public/secret key pair from the \textit{Gen algorithm}.
Consequently, banknotes can be transacted without relying on the outcomes of transactions involving other banknotes.

\subsection{Transitivity}
Our model necessitates the exchange of banknotes in a bilateral manner, meaning it involves the participation of two interconnected non-local Money Service Businesses (MSBs) representing Wallets. While it does not completely fulfill the local (offline) transitivity requirement, it ensures the fulfillment of other essential properties such as provenance/legitimacy, independence, and counterfeit detection imposed by transitivity.

\subsubsection{Confidentiality}
In all transactions, coordination between the payer and the payee is facilitated by their respective MSBs. Furthermore, since the MSB is the funds custodian, it requires the source and destination information as well as the amount to complete the transaction. Therefore, in both the intra-MSB and inter-MSB cases, the MSBs are able to observe the sender, the recipient and the transaction amount. Despite the need for network connectivity between Wallets and MSBs, as well as between MSBs of the payer and payee, privacy from the issuing authority is maintained. Information exchanged between the Wallet and the MSB consists solely of notifications devoid of any sensitive payment transaction details. 

Confidentiality of communications between vaults are protected against third-party entities as the message exchange utilizes a quantum network. Additionally, for an inter-MSB transaction, the settlement involves the physical quantum destruction of the banknote by the MSB, utilizing the \textit{GenCert algorithm} and the creation of a new banknote through an interactive process between the MSB and the Issuing Authority (i.e., the \textit{banknote minting/acquisition process}).  

In both cases, the MSB must record Wallet credentials associated to the banknote, as our model necessitates the MSB to store quantum banknotes on behalf of the Wallet. This is crucial for identifying the banknote's owner and, consequently, the funds associated with the Wallet. However, this is an MSB internal mechanism that is independent of the payment transaction. Neither the Wallet nor the Issuing Authority needs to be informed of this association during an inter-MSB transfer. From each MSB's perspective, their user is identified while the other user is pseudonymous (account number). Transactions are linkable as transaction behaviour is tied to specific quantum vaults. Since the Issuing Authority is not involved in transactions, it has zero insight into the behaviour of individual vaults. Minting and destruction requests to the Issuing Authority are at the MSB layer and can be batched across multiple user requests. In summary, the privacy mode is similar to the banking-financial system, where the issuing authority has minimal insight into transactions but MSBs, as active participants in the funds transfer mechanism, can observe user behaviour. 

It is noted that additional classical techniques can be introduced to enhance system privacy. Non-registered use could decouple vault behaviour from individuals. Similarly, the use of anonymous credentials for registration could create a separation of concerns between the authority that issues credentials and the MSBs that act as custodians of the vaults.

\section{Conclusions and Future Work}
\label{sec:Conclusions and Future Work}
We present a model for a \emph{quantum} currency (QC), utilizing a \textit{Public-key semi-quantum money scheme}. Notably, the issuing authority and wallets operate in a purely classical manner, with quantum hardware limited to intermediary \textit{quantum vaults}, designated as quantum-enabled Money Services Businesses (MSBs).
This innovative model effectively addresses the issue of double-spending and successfully fulfills the majority of the Security and Privacy Properties outlined in the context of our research.

This work highlights a trust concern in bilateral transactions, specifically offline payment transactions and quantum banknote transfers, where the  recipient (Wallet or MSB) has the unilateral power to reject the transferred banknote. We propose that this may be addressed via verifiable means for delegated quantum computation \cite{BFK09,Bro18,CGJV19}, and suggest it as future work.

\bibliographystyle{IEEEtran}
\input{main.bbl}


\end{document}

%% file: Styles.tikzstyles
\tikzstyle{new style 0}=[fill={rgb,255: red,191; green,191; blue,191}, draw=white, shape=circle]

\tikzstyle{black double direction arrow}=[draw=black, fill=black, <->]
\tikzstyle{light grey edge}=[-, fill={rgb,255: red,191; green,191; blue,191}, draw={rgb,255: red,191; green,191; blue,191}]
\tikzstyle{light grey double direction arrow}=[draw={rgb,255: red,191; green,191; blue,191}, fill={rgb,255: red,191; green,191; blue,191}, <->]
\tikzstyle{black one-direction arrow East}=[fill=black, ->]
\tikzstyle{black one-direction arrow - West}=[fill=black, <-]
\tikzstyle{black}=[-, fill=black]

%% file: QuantumVaultSystem.tikz
\begin{tikzpicture}
	\begin{pgfonlayer}{nodelayer}
		\node [style=none] (0) at (-4, 6) {};
		\node [style=none] (1) at (4, 6) {};
		\node [style=none] (2) at (4, 4) {};
		\node [style=none] (3) at (-4, 4) {};
		\node [style=none] (4) at (0, 5) {};
		\node [style=none] (5) at (0, 5) {Issuance/Redemption};
		\node [style=none] (6) at (-11, 0) {};
		\node [style=none] (7) at (-5, 0) {};
		\node [style=none] (8) at (-5, -2) {};
		\node [style=none] (9) at (-11, -2) {};
		\node [style=none] (10) at (-3, 0) {};
		\node [style=none] (11) at (-3, -2) {};
		\node [style=none] (12) at (3, -2) {};
		\node [style=none] (13) at (3, 0) {};
		\node [style=none] (14) at (5, 0) {};
		\node [style=none] (15) at (5, -2) {};
		\node [style=none] (16) at (11, -2) {};
		\node [style=none] (17) at (11, 0) {};
		\node [style=none] (20) at (0, -8) {};
		\node [style=none] (21) at (8, -8) {};
		\node [style=none] (22) at (-8.5, -1) {Quantum Vault};
		\node [style=none] (23) at (0, -1) {Quantum Vault};
		\node [style=none] (24) at (8, -1) {Quantum Vault};
		\node [style=new style 0] (25) at (-10, -8) {A};
		\node [style=new style 0] (26) at (-6, -8) {B};
		\node [style=new style 0] (27) at (0, -8) {C};
		\node [style=new style 0] (28) at (8, -8) {D};
		\node [style=none] (29) at (0, 4) {};
		\node [style=none] (34) at (-12, 1) {};
		\node [style=none] (35) at (-12, -3) {};
		\node [style=none] (36) at (12, -3) {};
		\node [style=none] (37) at (12, 1) {};
		\node [style=none] (38) at (-12, -7) {};
		\node [style=none] (39) at (-12, -9) {};
		\node [style=none] (40) at (12, -7) {};
		\node [style=none] (41) at (12, -9) {};
		\node [style=none] (42) at (-8, -2) {};
		\node [style=none] (43) at (0, -2) {};
		\node [style=none] (44) at (8, -2) {};
		\node [style=none] (50) at (0, -4.75) {Classical authorized request};
		\node [style=none] (51) at (-5, -1) {};
		\node [style=none] (52) at (-3, -1) {};
		\node [style=none] (53) at (3, -1) {};
		\node [style=none] (54) at (5, -1) {};
		\node [style=none] (57) at (-0.5, 6.5) {Issuing Authority Layer};
		\node [style=none] (59) at (-9.5, 1.5) {Intermediary Layer};
		\node [style=none] (60) at (-8.25, -6.5) {End-User Layer (Wallet)};
		\node [style=none] (61) at (-10, 0.25) {MSB 1};
		\node [style=none] (62) at (-2, 0.25) {MSB 2};
		\node [style=none] (63) at (6, 0.25) {MSB 3};
		\node [style=none] (64) at (-8, 0) {};
		\node [style=none] (65) at (0, 0) {};
		\node [style=none] (66) at (8, 0) {};
	\end{pgfonlayer}
	\begin{pgfonlayer}{edgelayer}
		\draw (0.center) to (1.center);
		\draw (1.center) to (2.center);
		\draw (2.center) to (3.center);
		\draw (3.center) to (0.center);
		\draw (6.center) to (7.center);
		\draw (7.center) to (8.center);
		\draw (8.center) to (9.center);
		\draw (9.center) to (6.center);
		\draw (10.center) to (11.center);
		\draw (11.center) to (12.center);
		\draw (12.center) to (13.center);
		\draw (13.center) to (10.center);
		\draw (14.center) to (15.center);
		\draw (15.center) to (16.center);
		\draw (16.center) to (17.center);
		\draw (17.center) to (14.center);
		\draw [style=light grey edge] (34.center) to (37.center);
		\draw [style=light grey edge] (37.center) to (36.center);
		\draw [style=light grey edge] (36.center) to (35.center);
		\draw [style=light grey edge] (35.center) to (34.center);
		\draw [style=light grey edge] (40.center) to (41.center);
		\draw [style=light grey edge] (41.center) to (39.center);
		\draw [style=light grey edge] (39.center) to (38.center);
		\draw [style=light grey edge] (38.center) to (40.center);
		\draw [style=black double direction arrow] (42.center) to (25);
		\draw [style=black double direction arrow] (42.center) to (26);
		\draw [style=black double direction arrow] (43.center) to (27);
		\draw [style=black double direction arrow] (44.center) to (28);
		\draw [style=light grey double direction arrow] (51.center) to (52.center);
		\draw [style=light grey double direction arrow] (53.center) to (54.center);
		\draw [style=black double direction arrow] (29.center) to (64.center);
		\draw [style=black double direction arrow] (29.center) to (65.center);
		\draw [style=black double direction arrow] (29.center) to (66.center);
	\end{pgfonlayer}
\end{tikzpicture}

%% file: OnDemandBanknoteMinting.tikz
\begin{tikzpicture}
	\begin{pgfonlayer}{nodelayer}
		\node [style=none] (0) at (-2, 4) {};
		\node [style=none] (1) at (-2, -18) {};
		\node [style=none] (2) at (12.5, 4) {};
		\node [style=none] (3) at (12.5, -18) {};
		\node [style=none] (8) at (-2, 4.5) {};
		\node [style=none] (9) at (12.5, 4.5) {};
		\node [style=none] (12) at (-2, 4.5) {Wallet};
		\node [style=none] (13) at (12.5, 4.5) {MSB};
		\node [style=none] (17) at (-2, 2) {};
		\node [style=none] (18) at (-2, 2) {};
		\node [style=new style 0] (19) at (-2, 2) {};
		\node [style=none] (24) at (12.5, 0) {};
		\node [style=none] (25) at (2.75, 2.25) {Request banknote};
		\node [style=none] (38) at (-2, -18) {};
		\node [style=none] (39) at (12.5, -16) {};
		\node [style=none] (41) at (-2, -16) {};
		\node [style=none] (43) at (2.75, -15.75) {Notify banknote status};
		\node [style=none] (45) at (9, 6) {};
		\node [style=none] (46) at (9, -19) {};
		\node [style=none] (48) at (16.25, 6) {};
		\node [style=none] (49) at (12.5, 6.25) {Quantum infrastructure};
		\node [style=none] (50) at (27, 4) {};
		\node [style=none] (51) at (27, -18) {};
		\node [style=none] (52) at (27, 4.5) {};
		\node [style=none] (53) at (27, 4.5) {Issuing Authority};
		\node [style=none] (75) at (16.25, -19) {};
		\node [style=none] (76) at (12.5, 2) {};
		\node [style=none] (77) at (27, 0) {};
		\node [style=none] (78) at (17.5, 0.25) {Request banknote};
		\node [style=none] (79) at (27, -2) {};
		\node [style=none] (80) at (27, -2) {Mint classical banknote};
		\node [style=none] (81) at (27, -4) {};
		\node [style=none] (82) at (12.5, -4) {};
		\node [style=none] (83) at (12.5, -6) {Mint quantum banknote};
		\node [style=none] (84) at (17.5, -3.75) {};
		\node [style=none] (85) at (17.5, -3.75) {Provide classical banknote};
		\node [style=none] (86) at (12.5, -8) {};
		\node [style=none] (87) at (27, -8) {};
		\node [style=none] (88) at (27, -10) {Validate classical cipher};
		\node [style=none] (89) at (27, -12) {};
		\node [style=none] (90) at (12.5, -12) {};
		\node [style=none] (91) at (12.5, -14) {Store banknote (physically)};
		\node [style=none] (93) at (-7.75, -8.75) {};
		\node [style=none] (95) at (19.75, -7.75) {Provide classical cipher of quantum banknote};
		\node [style=none] (96) at (17.5, -11.75) {Notify banknote validation status};
	\end{pgfonlayer}
	\begin{pgfonlayer}{edgelayer}
		\draw (0.center) to (1.center);
		\draw (2.center) to (3.center);
		\draw [style=black one-direction arrow - West] (41.center) to (39.center);
		\draw [style=light grey edge] (45.center) to (48.center);
		\draw [style=light grey edge] (46.center) to (45.center);
		\draw [style=black] (50.center) to (51.center);
		\draw [style=light grey edge] (48.center) to (75.center);
		\draw [style=light grey edge] (75.center) to (46.center);
		\draw [style=black one-direction arrow East] (19) to (76.center);
		\draw [style=black one-direction arrow East] (24.center) to (77.center);
		\draw [style=black one-direction arrow - West] (82.center) to (81.center);
		\draw [style=black one-direction arrow East] (86.center) to (87.center);
		\draw [style=black one-direction arrow East] (89.center) to (90.center);
	\end{pgfonlayer}
\end{tikzpicture}

%% file: QuantumBanknoteTransferInterMSB.tikz
\begin{tikzpicture}
	\begin{pgfonlayer}{nodelayer}
		\node [style=none] (1) at (-11.25, 4) {};
		\node [style=none] (2) at (-11.25, -12) {};
		\node [style=none] (3) at (-1, 4) {};
		\node [style=none] (4) at (-1, -12) {};
		\node [style=none] (5) at (9, 4) {};
		\node [style=none] (6) at (9, -12) {};
		\node [style=none] (7) at (19.25, 4) {};
		\node [style=none] (8) at (19.25, -12) {};
		\node [style=none] (9) at (-11.25, 4.5) {};
		\node [style=none] (10) at (-1, 4.5) {};
		\node [style=none] (11) at (9, 4.5) {};
		\node [style=none] (12) at (19.25, 4.5) {};
		\node [style=none] (13) at (-11.25, 4.5) {Wallet (Payer)};
		\node [style=none] (14) at (-1, 4.5) {MSB (P)};
		\node [style=none] (15) at (9, 4.5) {MSB (R)};
		\node [style=none] (16) at (19.25, 4.5) {};
		\node [style=none] (17) at (19.25, 4.5) {Wallet (Receiver)};
		\node [style=none] (18) at (-11.25, 2) {};
		\node [style=none] (19) at (-11.25, 2) {};
		\node [style=new style 0] (20) at (-11.25, 2) {};
		\node [style=none] (21) at (19.25, 2) {};
		\node [style=none] (22) at (4, 2.25) {};
		\node [style=none] (23) at (4, 2.25) {Agree on a payment transaction};
		\node [style=none] (24) at (-11.25, 0) {};
		\node [style=none] (25) at (-1, 0) {};
		\node [style=none] (27) at (-6, 0.25) {Request banknote transfer};
		\node [style=none] (30) at (-1, -2) {};
		\node [style=none] (31) at (9, -2) {};
		\node [style=none] (32) at (4, -1.75) {};
		\node [style=none] (33) at (4, -1.75) {Transfer banknote (physically)};
		\node [style=none] (34) at (9, -4) {};
		\node [style=none] (35) at (9, -4) {Verify banknote};
		\node [style=none] (36) at (9, -6) {};
		\node [style=none] (37) at (-1, -6) {};
		\node [style=none] (38) at (3.75, -5.75) {};
		\node [style=none] (39) at (3.75, -5.75) {Accept or Reject banknote};
		\node [style=none] (41) at (9, -8) {};
		\node [style=none] (42) at (9, -8) {Store banknote (physically)};
		\node [style=none] (43) at (-11.25, -12) {};
		\node [style=none] (44) at (-1, -10) {};
		\node [style=none] (45) at (9, -10) {};
		\node [style=none] (46) at (-11.25, -10) {};
		\node [style=none] (47) at (19.25, -10) {};
		\node [style=none] (48) at (-6.25, -9.75) {Notify payment transaction status};
		\node [style=none] (49) at (14, -9.75) {Notify payment transaction status};
		\node [style=none] (51) at (-3.5, 6) {};
		\node [style=none] (52) at (-3.5, -13) {};
		\node [style=none] (53) at (11.5, -13) {};
		\node [style=none] (54) at (11.5, 6) {};
		\node [style=none] (56) at (4.25, 6.25) {Quantum infrastructure};
	\end{pgfonlayer}
	\begin{pgfonlayer}{edgelayer}
		\draw (1.center) to (2.center);
		\draw (3.center) to (4.center);
		\draw (5.center) to (6.center);
		\draw (7.center) to (8.center);
		\draw [style=black double direction arrow, in=180, out=0] (20) to (21.center);
		\draw [style=black one-direction arrow - West] (37.center) to (36.center);
		\draw [style=black one-direction arrow - West] (46.center) to (44.center);
		\draw [style=black one-direction arrow East] (45.center) to (47.center);
		\draw [style=black one-direction arrow East] (30.center) to (31.center);
		\draw [style=black one-direction arrow East] (24.center) to (25.center);
		\draw [style=light grey edge] (51.center) to (54.center);
		\draw [style=light grey edge] (54.center) to (53.center);
		\draw [style=light grey edge] (53.center) to (52.center);
		\draw [style=light grey edge] (52.center) to (51.center);
	\end{pgfonlayer}
\end{tikzpicture}

%% file: OnlinePaymentTransactionInterMSB.tikz
\begin{tikzpicture}
	\begin{pgfonlayer}{nodelayer}
		\node [style=none] (0) at (-0.75, 4) {};
		\node [style=none] (1) at (-0.75, -22) {};
		\node [style=none] (2) at (9.25, 4) {};
		\node [style=none] (3) at (9.25, -22) {};
		\node [style=none] (4) at (23.75, 4) {};
		\node [style=none] (5) at (23.75, -22) {};
		\node [style=none] (6) at (47, 4) {};
		\node [style=none] (7) at (47, -22) {};
		\node [style=none] (8) at (-0.75, 4.5) {};
		\node [style=none] (9) at (9.25, 4.5) {};
		\node [style=none] (10) at (23.75, 4.5) {};
		\node [style=none] (11) at (47, 4.5) {};
		\node [style=none] (12) at (-0.75, 4.5) {Wallet (Payer)};
		\node [style=none] (13) at (9.25, 4.5) {MSB (P)};
		\node [style=none] (14) at (23.75, 4.5) {MSB (R)};
		\node [style=none] (15) at (47, 4.5) {};
		\node [style=none] (16) at (47, 4.5) {Wallet (Receiver)};
		\node [style=none] (17) at (-0.75, 2) {};
		\node [style=none] (18) at (-0.75, 2) {};
		\node [style=new style 0] (19) at (-0.75, 2) {};
		\node [style=none] (20) at (47, 2) {};
		\node [style=none] (22) at (23.75, 2.25) {Agree on a payment transaction};
		\node [style=none] (23) at (-0.75, 0) {};
		\node [style=none] (24) at (9.25, 0) {};
		\node [style=none] (25) at (4, 0.25) {Request banknote transfer};
		\node [style=none] (26) at (9.25, -4) {};
		\node [style=none] (27) at (23.75, -4) {};
		\node [style=none] (28) at (16.75, -3.75) {};
		\node [style=none] (29) at (16.75, -3.75) {Request banknote transfer};
		\node [style=none] (30) at (37.25, -12) {};
		\node [style=none] (31) at (37.25, -12) {Verify banknote destruction};
		\node [style=none] (37) at (23.75, -16) {Store banknote (physically)};
		\node [style=none] (38) at (-0.75, -22) {};
		\node [style=none] (39) at (9.25, -20) {};
		\node [style=none] (40) at (23.75, -20) {};
		\node [style=none] (41) at (-0.75, -20) {};
		\node [style=none] (42) at (47, -20) {};
		\node [style=none] (43) at (4.25, -19.75) {Notify payment transaction status};
		\node [style=none] (44) at (30.75, -19.75) {Notify payment transaction status};
		\node [style=none] (45) at (8, 6) {};
		\node [style=none] (46) at (8, -23) {};
		\node [style=none] (47) at (25, -23) {};
		\node [style=none] (48) at (25, 6) {};
		\node [style=none] (49) at (16, 6.25) {Quantum infrastructure};
		\node [style=none] (50) at (37.25, 4) {};
		\node [style=none] (51) at (37.25, -22) {};
		\node [style=none] (52) at (37.25, 4.5) {};
		\node [style=none] (53) at (37.25, 4.5) {Issuing Authority};
		\node [style=none] (54) at (9.25, -2) {};
		\node [style=none] (55) at (9.25, -2) {Destroy Banknote (physically)};
		\node [style=none] (56) at (9.25, -6) {};
		\node [style=none] (57) at (23.75, -6) {};
		\node [style=none] (58) at (16.5, -5.75) {};
		\node [style=none] (59) at (16.5, -5.75) {Transfer certificate of banknote destruction};
		\node [style=none] (63) at (23.75, -8) {};
		\node [style=none] (64) at (37.25, -8) {};
		\node [style=none] (65) at (30.75, -7.75) {};
		\node [style=none] (66) at (30.75, -7.75) {Request banknote (minting)};
		\node [style=none] (67) at (23.75, -10) {};
		\node [style=none] (68) at (37.25, -10) {};
		\node [style=none] (69) at (30.75, -9.75) {Transfer certificate of banknote destruction};
		\node [style=none] (70) at (23.75, -14) {};
		\node [style=none] (71) at (37.25, -14) {};
		\node [style=none] (72) at (30.75, -13.75) {Mint banknote};
		\node [style=none] (73) at (23.75, -18) {};
		\node [style=none] (74) at (9.25, -18) {};
		\node [style=none] (75) at (16.25, -17.75) {};
		\node [style=none] (76) at (16.25, -17.75) {Notify banknote transfer status};
	\end{pgfonlayer}
	\begin{pgfonlayer}{edgelayer}
		\draw (0.center) to (1.center);
		\draw (2.center) to (3.center);
		\draw (4.center) to (5.center);
		\draw (6.center) to (7.center);
		\draw [style=black double direction arrow, ] (19) to (20.center);
  
		\draw [style=black one-direction arrow - West] (41.center) to (39.center);
		\draw [style=black one-direction arrow East] (40.center) to (42.center);
		\draw [style=black one-direction arrow East] (26.center) to (27.center);
		\draw [style=black one-direction arrow East] (23.center) to (24.center);
		\draw [style=light grey edge] (45.center) to (48.center);
		\draw [style=light grey edge] (48.center) to (47.center);
		\draw [style=light grey edge] (47.center) to (46.center);
		\draw [style=light grey edge] (46.center) to (45.center);
		\draw [style=black] (50.center) to (51.center);
		\draw [style=black one-direction arrow East] (56.center) to (57.center);
		\draw [style=black one-direction arrow East] (63.center) to (64.center);
		\draw [style=black one-direction arrow East] (67.center) to (68.center);
		\draw [style=black double direction arrow] (70.center) to (71.center);
		\draw [style=black one-direction arrow - West] (74.center) to (73.center);
	\end{pgfonlayer}
\end{tikzpicture}

%% file: main.bbl
\makeatletter\@ifundefined{url}{\newcommand{\url}[1]{\texttt{#1}}}{}\@ifundefined{href}{\newcommand{\href}[2]{\texttt{#2}}}{}\@ifundefined{mathbb}{\newcommand{\mathbb}[1]{#1}}{}\makeatother